\documentclass{aa}
\usepackage{graphicx}
\usepackage{longtable}

\def\gtrsim{\mathrel{\hbox{\rlap{\hbox{\lower4pt\hbox{$\sim$}}}\hbox{$>$}}}}
\def\ltsim{\mathrel{\hbox{\rlap{\hbox{\lower4pt\hbox{$\sim$}}}\hbox{$<$}}}}

\begin{document}

\title{No detection of large-scale magnetic fields at the surfaces of Am and HgMn stars\thanks{Based on data obtained using the T\'elescope Bernard Lyot at Observatoire du Pic du Midi, CNRS/INSU and Universit\'e de Toulouse, France.}}

\author{M. Auri\`ere\inst{1}, G.A. Wade\inst{2}, F. Ligni\`eres\inst{1}, A. Hui-Bon-Hoa\inst{1},   J.D. Landstreet\inst{3,4}, I. Iliev\inst{5}, J.-F. Donati\inst{1}, P. Petit\inst{1},  T. Roudier \inst{1},  S. Th\'eado \inst{1} }
\offprints{M. Auri\`ere, {\tt michel.auriere@ast.obs-mip.fr}}
\institute{Laboratoire d'Astrophysique de Toulouse-Tarbes, CNRS, Universit\'e de Toulouse, 57 Avenue d'Azereix, 65008 Tarbes, France\\
\email{michel.auriere@ast.obs-mip.fr}
\email{ligniere@ast.obs-mip.fr}
\email{alain.hui@ast.obs-mip.fr}
\email{donati@ast.obs-mip.fr}
\email{petit@ast.obs-mip.fr}
\email{roudier@ast.obs-mip.fr}
\email{stheado@ast.obs-mip.fr}
 \and
Department of Physics, Royal Military College of Canada,
  PO Box 17000, Station 'Forces', Kingston, Ontario, Canada K7K 4B4\\
\email{gregg.wade@rmc.ca}
\and
 Department of Physics \& Astronomy, The University of Western Ontario, London, Ontario, Canada, N6A 3K7\\
\email{jlandstr@uwo.ca}
\and
Armagh Observatory, College Hill, Armagh, Northern Ireland BT61 9DG\\
\email{jls@arm.ac.uk}
\and 
Institute of Astronomy, Bulgarian Academy of Sciences, 72 Tsarigradsko shose, 1784 Sofia, Bulgaria\\
\email{iliani@astro.bas.bg}}

   \date{Received ??; accepted ??}

\abstract {}
{We investigate the magnetic dichotomy between Ap/Bp and other A-type stars by carrying out a deep spectropolarimetric study of Am and HgMn stars.}
{Using  the NARVAL spectropolarimeter at the T\'elescope Bernard 
Lyot (Observatoire du Pic du Midi, France), we obtained high-resolution circular polarisation spectroscopy of 12 Am stars and 3 HgMn stars.}
{Using Least Squares Deconvolution (LSD),  no magnetic field is detected  in any of the 15 observed stars. Uncertaintiies as low as 0.3 G (respectively 1 G) have been reached for surface-averaged longitudinal magnetic field measurements for Am (respectively HgMn) stars. }{Associated with the results obtained previously for Ap/Bp stars, our study confirms the existence of a magnetic dichotomy among A-type stars. Our data demonstrate that there is at least one order of magnitude difference in field strength  between Zeeman detected stars (Ap/Bp stars) and non Zeeman detected stars (Am and HgMn stars). This result confirms that the spectroscopically-defined Ap/Bp stars are the only A-type stars harbouring detectable large-scale surface magnetic fields. }{}
\keywords{stars: chemically peculiar -- stars: magnetic field}
\authorrunning{M. Auri\`ere et al.}
\titlerunning{ Magnetic fields of Am and HgMn stars}

\maketitle

\section{Introduction}

The spectroscopically-selected ``magnetic Ap/Bp stars'' (hereafter Ap/Bp stars), corresponding to about 5\% of main sequence (MS) A and B stars (Wolff 1968), are known to host relatively strong, ordered magnetic fields. On the other hand, the remaining 95\% of MS stars at these spectral types appear to have no detectable magnetic field (with the exception of the very small magnetic field recently detected on Vega by Ligni\`eres et al. 2009, Petit et al. 2010). This is the so-called magnetic dichotomy. Using the MuSiCoS and NARVAL spectropolarimeters, Auri\`ere et al. (2007) studied the weak part of the magnetic field distribution of Ap/Bp stars. They found, as had previously been assumed, that all confidently spectroscopically-classified Ap/Bp stars, when observed with sufficient precision and tenacity, show evidence for organised magnetic fields with model dipole polar strength stronger than about 300~G. However, demonstrating the existence of a magnetic field dichotomy relies not only on establishing the universal presence of large scale fields in Ap/Bp stars, but also showing confidently that no such fields are detectable in the non-Ap/Bp stars. The most recent 
sensitive surveys of apparently non magnetic A and B stars have led to non-detection of magnetic fields at the level of a few tens of G (Shorlin et al. 2002, Bagnulo et al. 2006). Shorlin et al. (2002) used the high-resolution  MuSiCoS spectropolarimeter to search for Stokes $V$ Zeeman signatures in spectra of 63 non-Ap/Bp intermediate-mass stars, finding no evidence of magnetic fields, with a median longitudinal field ($B_\ell$) formal error of just 22 G. Bagnulo et al. (2006) used the low-resolution FORS1 spectropolarimeter to measure magnetic fields of a large sample of intermediate-mass stars in open clusters. In their sample of 138 non-Ap/Bp stars, no magnetic field was detected, with a median longitudinal field error bar of 136 G. To refine our understanding of the dichotomy, using the possibilities of new instruments (Donati \& Landstreet 2009), in this study we employ NARVAL to observe bright slow rotators among the Am and HgMn stars previously studied by Shorlin et al (2002). 
In the following, we will describe our observations (Sect. 2) and our results for each category of stars (Sect. 3), then give our discussion of the magnetic dichotomy and our conclusions.

\section{Observations and reduction}

\subsection{Observations with NARVAL}

The observations  took place in March 2007, at the 2-m
T\'{e}lescope Bernard Lyot (TBL) of Pic du 
Midi Observatory with the NARVAL high-resolution spectropolarimeter (Auri\`ere 2003). In operation since December 2006, NARVAL is a copy of ESPaDOnS 
installed at CFHT at the end of 2004 
(Donati et al. 2006). NARVAL is a fiber--fed echelle spectropolarimeter 
with which the whole spectrum from 370 nm to 1000 nm is recorded in each 
exposure. The  40 grating orders are aligned on the CCD frame using two cross-disperser prisms.
NARVAL was used in polarimetric mode with a spectral resolution of 
about 65000. Stokes $I$ (unpolarised) and Stokes $V$ (circular polarisation) 
parameters were obtained by means of 4 sub-exposures between which the 
retarders (Fresnel rhombs) were rotated in order to exchange the beams 
in the whole instrument and to reduce spurious polarization signatures. 
We aimed to get long exposures, up to 6400s, on our bright targets in order to be able to detect ultra-weak or complex magnetic fields. In order to avoid saturation of the CCD we made short sub exposures (e.g. 4 or 8 second subexposures for each Stokes $V$ series in the case of Sirius).

\subsection{Reduction and magnetic field detection}

 During the technical tests and science demonstration time, magnetic and non magnetic stars were observed 
which showed that NARVAL works properly and is 30 times more efficient than the previous instrument, MuSiCoS 
(Baudrand \& B\"ohm 1992, Donati et al. 1999), which was used by Shorlin et al. (2002). Since then, a great number of new results have been obtained that confirm the high scientific efficiency of ESPaDOnS and NARVAL (e.g. in Donati \& Landstreet 2009).
The extraction of the spectra was done using Libre-ESpRIT (Donati et al. 1997), 
a fully automatic reduction package installed at the TBL. In 
order to carry out the Zeeman analysis, Least-Squares Deconvolution analysis 
(LSD, Donati et al. 1997) was applied to all observations. We used line masks with solar abundances, $\log g$ = 4, temperatures close to the  values given by Shorlin et al. (2002; see our Table 1), and included lines with a central depth greater than 10\% of the continuum. For our sample, this method enabled us to average from about 500 (highest temperature HgMn star) to about 5000 (coolest Am star) lines and to obtain Stokes $V$ profiles with signal-to-noise ratio (S/N) increased by a factor of about 10 to 40. We performed a statistical test for the detection of Stokes $V$ Zeeman signatures: the reduced $\chi^2$ statistic is computed for zero signal in the Stokes $V$ profile, both inside and outside the spectral line (Donati et al. 1997). The statistics are then converted into detection probabilities (false alarm probability). Also included in the output are ``diagnostic null'' spectra $N$ (combinations of sub-exposures in which real $V$ signatures should cancel out), which are in principle featureless, and therefore serve to diagnose the presence of spurious contributions to the Stokes $V$ spectra.
We then computed the longitudinal magnetic field $B_{l}$ in G, using the 
first-order moment method adapted to LSD profiles (Rees and Semel 1979, Donati et al. 1997, Wade et al. 2000). The integration range used to compute $B_{l}$ corresponds to the first and last point in the Stokes $I$ profile for which the flux was lower than 15$\%$ of the maximum depth, except for the SB2 stars for which it was optimized manually. 

For a few selected stars we constructed line masks that matched the stellar spectrum in detail, by modifying individual line depths in the mask, using data provided by VALD (Kupka et al. 1999). While these custom masks naturally provided a better representation of the Stokes $I$ and $V$ spectra, they did not result in any change in the detection diagnosis, or any significant improvement in the longitudinal field upper limit. As a consequence, all results presented here correspond to solar abundance line masks.

Finally, we measured for each star (generally the primary) the radial velocity $RV$ from the averaged LSD Stokes $I$ profile, using a gaussian fit. The long term stability of NARVAL is about 30~m/s (e.g. Auri\`ere et al. 2009a) but the absolute uncertainty of individual measurements relative to the local standard of rest is about 1~km~s$^{-1}$.

Table 1 gives for each star its $V$ magnitude, spectral class, mask temperature used, $v \sin i$, and, for each observation, the date, HJD (corresponding to the $RV$ measurement),  number of exposures and total exposure time, $RV$, and the inferred longitudinal magnetic field with its standard error in G.

\begin{table*}[t]

\caption{Summary of observations. } 
\begin{tabular}{llccccccccccc}
\\
\hline
\hline
ID             & HD    & $V$& Spec. & Mask & $v \sin i$ & Date     &HJD      &\# & Exp.& $RV        $ & $B_\ell$ & $\sigma$\\
               &       &mag.&       &  K          & ~km~s$^{-1}$&          &2450000 +&   &        s &~km~s$^{-1}$ & G     & G       \\
\hline
Am Stars       &       &    &       &             &             &          &         &         &            &         &       &         \\
\hline
Sirius         &  48915&  -1.47 &    A1m& 10000&16.5 (2)    & 12Mar07&4172.33& 32    &  1024    & -7.3       & -0.10 & 0.32    \\ 
$\alpha$ Gem B &  60178&    2.9     &A2m& 9000&20 (1)     & 13Mar07&4173.41& 4     &  2800    & -16.3        & -0.40 & 0.79    \\ 
15 UMa         &  78209&    4.5     &F3m& 7500&38 (1)     & 14Mar07&4174.43& 1     &  3200    & +0.7        & -1.74 & 2.17    \\ 
$\tau$ UMa     &  78362&    4.6     &F3m& 7250&11.3 (1)   & 11Mar07&4171.39& 1     &  3200    & -8.7       & -0.6  & 0.54    \\
$\lambda$ UMa  &  89021&    3.4     &A2m& 9000&53 (3)     & 11Mar07&4171.51& 3     &  3600    & +19.9       &  2.77 & 2.91    \\
$\beta$ UMa    &  95418&    2.3     &A1V& 9500&48 (3)     & 14Mar07&4174.50& 6     &  2880    & -12.0      & -3.03 & 2.93    \\
$\theta$ Leo   &  97633&    3.3     &A2V& 9250&23.5 (1)   & 12Mar07&4172.53& 3     &  3600    & +7.4        &  1.60 & 1.39    \\
32 Vir         & 110951&    5.2     &F0IVm& 7250&19.3 (1)   & 12Mar07&4172.58& 1     &  3600    & -39.3      &  2.15 & 1.62    \\
               &       &            &    &   &          & 13Mar07&4173.58& 1     &  3600    & -45.0      & -1.27 & 1.53    \\
               &       &           &   &         &      &02Apr08&4559.49& 1     &  3600    & -57.5      & -2.53 & 1.98    \\
$\lambda$ Vir  & 125337&  4.5 &    A2m  & 9500 &   & 13Mar07&4173.63& 2     &  3200    &            & -4.47 & 2.88    \\
 Comp. A       &       &       &   &    & 36 (4)     & 13Mar07&4173.63& 2     &  3200    & +11.5       & -1.20 & 2.69    \\
 Comp. B       &       &       &    &   & 10 (4)     & 13Mar07&4173.63& 2     &  3200    & -27.9      & -3.12 & 1.41    \\
22 Boo         & 126661&    5.4&    F0m& 8000     & 36 (1)     & 13Mar07&4173.68& 1     &  3200    & -27.6      & -1.44 & 2.18    \\
               & 141675&    5.8&    A3m& 8000     & 33 (1)     & 14Mar07&4174.66& 1     &  3600    & -0.8       & -4.91 & 3.11    \\
$\epsilon$ Ser & 141795&    3.7&    A2m& 8500     & 33.5 (1)   & 12Mar07&4172.65& 2     &  3200    & -10.0      & -0.79 & 1.40    \\ 
\hline
HgMn           &       &       &       &         &        &       &       &          &         &       &         \\
\hline 
$\kappa$ Cnc   &  78316&    5.2&     B8& 13000   & 6.8 (2)    & 13Mar07&4173.46& 1     &  3200    & +76.2       &  -1.05& 3.12    \\
$\iota$ CrB    & 143807&    4.9&     A0& 10500   & 1.0 (5)    & 11Mar07&4171.67& 2     &  6400    & -19.6      & 0.07  & 1.34    \\ 
$\phi$  Her    & 145389&    4.2&     B9& 11000   & 8.0 (6)   & 12Mar07&4172.70& 3     &  4800    & -15.4      & -1.76 & 1.99    \\ 

\hline
\hline
\end{tabular}
Note: Individual columns show target ID, HD number, spectral class, mask temperature, $v \sin i$ (number of reference given at the end of the note), date observed, associated HJD, number of exposures acquired, total exposure time, $RV$, longitudinal field $B_\ell$ and its uncertainty $\sigma$. References for $v \sin i$: (1) Shorlin et al. 2002; (2) Landstreet et al. 2009; (3) Fekel 2003; (4) Zaho et al. 2007; (5) Dubaj et al. 2005; (6) Zavala et al. 2007.

\end{table*}

\section{Results of the present survey}

\subsection{The sample}

Our aim is to search for magnetic fields on non Ap/Bp A-type stars in order to establish definitively the gap of the magnetic dichotomy.  Shorlin et al. (2002) showed the great influence of the value of $v \sin i$ on the sensitivity of a magnetic survey using high-resolution spectropolarimetry. In order to reduce the errors in our survey, we choose here to observe the most promising objects already observed by Shorlin et al. (2002). 

Am stars are frequently found in close binaries (Abt \& Levy 1985), likely because tidal interactions in such systems slow stellar rotation and thereby reduce rotational mixing. This is also the case for HgMn stars (Ryabchikova 1998). This property does not hamper our study, but the interesting cases of the SB2 stars 32 Vir and $\lambda$ Vir are discussed in detail in Sect.3.2. No Zeeman detection was obtained for any of our sample stars, since false alarm probability was always greater than 10$^{-3}$, apart from the case of 32 Vir which is discussed in Sect. 3.2. The results are discussed further for Am and HgMn stars in Sect. 3.2 and 3.3 respectively.

\subsection{Am stars}

Am stars are cool A-type stars that can be considered as ''ordinary'' slowly rotating A-stars (Takeda et al. 2008). A large number of Am stars deserve a sensitive magnetic survey with NARVAL; we observed 12 of them, among them the bright star Sirius. Our main selection criterion for the stars observed was low $v \sin i$, which we required to be smaller than 50~km~s$^{-1}$, and is often much smaller. 
Our sample stars are generally on the main sequence, but two of them, 32 Vir and 22 Boo, have already left it. The main result of our study (no Zeeman detection and very low upper limits for a possible surface-averaged longitudinal magnetic field) is visible on Table 1, but we make comments about some stars below.

Sirius:
 Besides being the brightest star in the sky after the Sun, Sirius is a hot Am star. Observing it enabled us to reach the highest precision obtained in our survey, namely 0.32 G for our $B_\ell$ determination. With 32 Stokes $V$ series, we got a total exposure of 1024 s.  Fig.1 shows the composite LSD profiles. The huge enlargement of Stokes $V$ and Stokes $N$ show that the amplitude of the noise is currently smaller than $10^{-5} I_c$ .  A kind of flat feature appears on Stokes $V$ profile at the position of the absorption line in the intensity profile. It is not significant with respect to the LSD detection statistical test. Splitting our spectra into two equal subsets show that this feature is more visible on  our second subset, and is probably due to noise. No magnetic field is therefore detected on Sirius and the corresponding $B_{l}$ value is $0.10 \pm 0.32$ G ($1 \sigma$).  Equally small or even smaller errors in Stokes $V$ profiles and $B_{l}$ measurements were obtained with NARVAL in the case of the normal A-star Vega (Ligni\` eres et al. 2009, Petit et al. 2010) and the red giant Pollux (Auri\`ere et al. 2009a), and sub-G magnetic fields could be detected at a significant level in these stars.

$\alpha$ Gem B (Castor B):
Castor is a multiple system composed of three visual stars, each of which is by itself a spectroscopic binary. Castor A and Castor C were out of the slit during the NARVAL observations. Castor A and B have been now resolved in X-rays and this observation shows that the late-type secondaries within each spectroscopic binary are the sites of the X-ray production (Stelzer \& Burwitz 2003). Our non-detection of a magnetic field confirms the absence of magnetic activity at the surface of the A-type star Castor B.

$\lambda$ UMa:
Our observations confirm that the 3 $\sigma$ detection ($66\pm 22$~G) of a magnetic field by Shorlin et al. (2002) is spurious, as suspected by those authors. We have improved the precision of field measurement by a factor greater than 7 with respect to the MuSiCoS result, though the corresponding error on $B_\ell$ is one of the least precise in this paper, 2.91 G, due to the relatively large $v \sin i$ of 50~km~s$^{-1}$.

$\theta$ Leo:
This star is considered to be a hot Am star (Smith 1974, Adelman 2004) and is a standard of radial velocity (Morse et al. 1991). Because of its moderate $v \sin i$ = 23.5~km~s$^{-1}$, the field measurement is one of the most accurate in the survey ($\sigma = $ 1.39 G) for early A-type stars. 

32 Vir:
32 Vir is a well known SB2 whose primary appears to be a $\rho$ Puppis-type $\delta$~Scuti star, i.e. an evolved pulsating Am star (Mitton \& Stickland 1979). Fig.2 shows the LSD Stokes profiles derived from a total integration of 3600s  obtained on 12 March 2007. The Stokes $I$ LSD profile easily resolves the two components, the Am star corresponding to the sharp line. In both the Stokes $V$ and null polarization $N$ profiles, a small signal is visible at the RV of the Am star, and the LSD statistics gives a false alarm probability smaller than 10$^{-3}$. Since this detection is obtained also, even more strongly, on the null polarization $N$ profile, it is suspected to be spurious. However weak magnetic fields in subgiant stars have been detected in the course of another magnetic survey with NARVAL (Auri\`ere et al. 2009b). We therefore observed 32 Vir again in the same conditions on 13 March 2007 and on 02 April 2008, and got the same result: a weak signal was again visible on Stokes $V$ and $N$ profiles. Now 32 Vir is both a binary star and a pulsating star (Lampens \& Boffin 2000). Bertiau (1957) derived an orbit with a period of 38.3 days and a semi amplitude of 48~km~s$^{-1}$. As a $\delta$ Scuti star, 32 Vir has a period of about 0.07 day (Bartolini et al. 1983, Kurtz et al. 1976). The RV amplitude variation due to pulsations is unknown, but could be similar to that observed for $\rho$ Puppis itself, i.e. 8.6~km~s$^{-1}$ (Mathias et al. 1997). These rapid $RV$ variations due to the binary and pulsating status of the star are expected to induce  shifts in $RV$ between the LSD profiles of the four (900 s) sub-exposures of up to about 1 km/s. Such large $RV$ shifts were actually measured on our data, which can lead to detection of spurious polarization signals (Donati et al. 1997). Because of the different time-lags between combinations of sub-exposures used for getting $N$ and Stokes $V$ profiles, the spurious signal is expected to be stronger on the former than on the latter profiles. This process is probably the reason for the signal observed on the 3 dates.

$\lambda$ Vir:
This star is a well-known double lined spectroscopic Am binary: both stars are very similar in chemical abundances but the primary component is broad-lined and the secondary  is sharp-lined (Zhao et al. 2007). Our NARVAL observations enabled us to resolve the two components on our LSD Stokes $V$ profiles, as already presented by Shorlin et al. (2002). In Table 1 we show that neither of the two components indicates a Zeeman detection and we have included individual $B_\ell$ measurements for each of the two components.

22 Boo:
22 Boo is considered to be an Am star which has already left the main sequence (Burkart et al. 1980, Bertet 1990). This is a particularly interesting object for a magnetic survey since a dynamo driven  magnetic field may appear during the subgiant phase (Auri\`ere et al. 2009b). However no Zeeman detection occur at a level of $\sigma = 2.18$~G for $B_\ell$.

\subsection{HgMn stars}

The HgMn stars are generally considered as having the most stable atmospheres among intermediate mass stars (Vauclair \& Vauclair, 1982). However, some binary HgMn stars have been shown to display spectroscopic variations (Adelman et al. 2002, Kochukhov et al. 2005, Hubrig et al. 2006a, Briquet et al. 2010). The non-uniform surface abundances invoked to explain these variations appear to evolve with time (Kochukhov et al. 2007). It has been proposed that they could host strong magnetic fields of peculiar topology (Hubrig et al. 2006b, 2008), and that such fields could be responsible for the surface structures. Wade et al. (2006) performed a sensitive magnetic study of the brightest HgMn star, $\alpha$ And, and placed a 3$\sigma$ upper limit of about 100 G on the possible presence of any undetected pure dipolar, quadrupolar or octupolar surface magnetic fields. Because of the rather large $v \sin i$ (52~km~s$^{-1}$), the 1$\sigma$ error bars reached 6 G at the smallest, even with ESPaDOnS. We have observed here with NARVAL 3 of the brightest of the HgMn stars having $v \sin i$ 5 times smaller than $\alpha$ And. The resulting uncertainties of $B_\ell$ are finally 2 to 4 times smaller than those obtained for $\alpha$ And.

$\kappa$ Cnc:
For this classical HgMn star (Z\"ochling \& Muthsam, 1987), our non detection with a 1 $\sigma$ error of 3 G for the longitudinal magnetic field confirms the result of Shorlin et al. 2002, that a strong surface magnetic field, as suggested by older observations, is not present.

$\iota$ CrB:
Observations of this star with the Gecko spectrograph attached to the Canada-France-Hawaii Telescope have resolved the two components of the spectroscopic binary (Dubaj et al. 2005). The $v \sin i$ of the HgMn component was measured to be only about  1~km~s$^{-1}$;  Shorlin et al. (2002) were only able to find an upper limit of $v \sin i$ $<$  10~km~s$^{-1}$.  Our measurement of  this star has the best precision obtained for the HgMn stars of our sample,  about 1~G. Figure 3  shows the Stokes $V$ and Stokes $I$ LSD profiles for $\iota$ CrB.

$\phi$ Her:
This star is a spectroscopic binary which has recently been resolved (Zavala et al. 2007) and for which the mass of the CP star has been refined (Torres 2007). No field is detected, with an longitudinal field uncertainty of about 2~G.

\begin{figure}[thp]
\includegraphics[width=8.cm,angle=0]{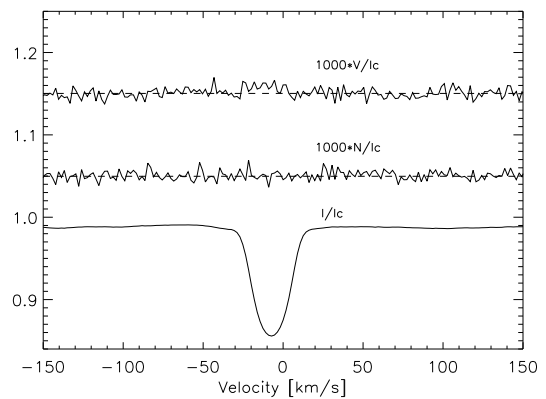} 
\caption{LSD profiles of the Am star Sirius as observed on 12 March 2007 with NARVAL. From bottom to top: Stokes $I$, zero polarization $N$ and Stokes $V$ profiles.  For display purposes, the profiles are shifted vertically and the Stokes $V$ and diagnostic $N$ profiles are expanded by a factor of 1000. The dashed line illustrates the zero level for the Stokes $V$ and null $N$ profiles. }
\label{f2}
\end{figure}

\begin{figure}[thp]
\includegraphics[width=8.cm,angle=0] {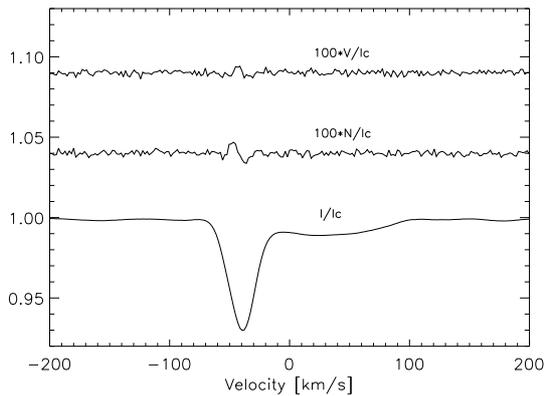} 
\caption{LSD profiles of the Am star 32 Vir. as observed on 12 March 2007 with NARVAL. From bottom to top: Stokes $I$, zero polarization $N$ and Stokes $V$ profiles.  For display purposes, the profiles are shifted vertically and the Stokes $V$ and diagnostic $N$ profiles are expanded by a factor of 100. }
\label{f3}
\end{figure}

\begin{figure}[thp]
\centerline{\includegraphics[width=8.cm,angle=0] {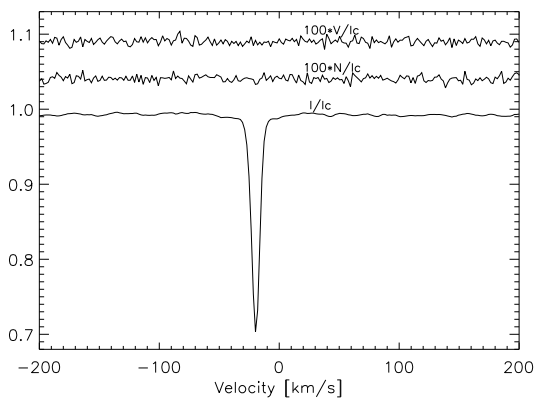}}
\caption{LSD profiles of the HgMn star $\iota$ CrB on 11 Mar 07 as observed with NARVAL  From bottom to top, Stokes $I$, zero polarization, and Stokes $V$ are presented. For display purposes, the profiles are shifted vertically, and  Stokes $V$  and diagnostic $N$ profiles are expanded by a factor of 100. }
\label{f2}
\end{figure}

\section{The magnetic dichotomy}

No Zeeman detection was obtained for any of the 15 stars of our sample, although we have achieved a precision improvement of more than one order of magnitude with respect to the work of Shorlin et al. (2002). Although we have obtained only one observation for the majority of the stars of our sample, the non-detection of significant Stokes $V$ signatures is a strong negative result because magnetic configurations can produce detectable  $V$ signatures through the line profile even for zero longitudinal magnetic field. The observation of the crossover effect requires non-negligible rotational Doppler broadening (Mathys 1995), but it could be observed in the case of HN And ($vsini$ = 2 km s$^{-1}$, Auri\`ere et al. 2007), and therefore could be observed on all our stars apart from $\iota$ CrB. Error bars in the range of 0.3 to 3 G have been obtained for our measurements of the surface-average longitudinal magnetic fields and can therefore be used to set upper limits of this component of the magnetic field of about 10 G (3 $\sigma$). Table 1 of Auri\`ere et al. (2007) shows that for the weak magnetic Ap/Bp stars,  $|B_{\ell}|^{max}$ is generally above 100 G, i.e. about 10 times stronger than the present upper limit. Therefore, a very significant gap of at least one order of magnitude is now established between upper limits of fields that might be present in non-detected Am/HgMn stars and the fields consistently detected in Ap/Bp stars. 

To interpret this result in term of magnetic intensity, some assumption has to be made for the magnetic topology. Taking into account the results of Auri\`ere et al. (2007) who deduced the existence of a threshold magnetic field of about 300 G at the surface of Ap/Bp stars, and for geometrical configurations similar to those observed in weakly Ap/Bp stars, large scale magnetic fields with dipole field strength greater than about  30 G are not present at the surface of Am and HgMn stars. Moreover, the high-resolution spectropolarimetric techniques used in this study have been shown to be sensitive to both the large-scale (e.g. Auri\`ere et al. 2009a) and smaller-scale (e.g. Petit et al. 2004) magnetic fields of active late-type stars. While there is certainly a spatial resolution limit to this sensitivity, the very high precision obtained in our survey definitely does not support previous reports of strong, complex fields in Am (Lanz \& Mathys 1993) and HgMn stars (Hubrig et al. 2006b, 2008).

   The report of a weak magnetic field (of about one G) in Vega (Ligni\`eres et al. 2009, Petit et al. 2010) is consistent with the  existence of a magnetic dichotomy in the A-type stars. The instability scenario of Auri\`ere et al. (2007) gives a possible explanation of this gap. The Ap/Bp stars are those for which the surface magnetic field is strong enough to resist to differential rotation and instabilities such as the Tayler instability (Spruit 1999). Conversely, stars with a large scale magnetic field of lower strength are subjected to instabilities that will strongly reduce their surface-averaged longitudinal field through cancellation effects. This dichotomy between stable and unstable large scale field configurations naturally leads to a gap in the values of the longitudinal fields.

\section{Conclusion}

Our limited survey of 15  A-type star of peculiarity types other than Ap/Bp shows that none of them appear to host a large scale magnetic field having a  surface-averaged longitudinal magnetic field of more than 3 G. Taken together with the result of Auri\`ere et al. (2007), who showed the existence of a magnetic threshold of about 300 G for dipole strength in Ap/Bp stars, this result confirms the existence of a magnetic dichotomy, and shows that it corresponds to a gap of more than one order of magnitude in field strength. In fact, up to now a magnetic field has been detected by spectropolarimetry for a non Ap/Bp star only in Vega, and the surface averaged-longitudinal magnetic field appears to be smaller than one G (Ligni\`eres et al. 2009, Petit et al. 2010). Our result can be simply explained by the instability scenario described in Auri\`ere et al. (2007).


\begin{acknowledgements}
 This research has made use of databases operated by CDS, Strasbourg, France, and of VALD (Vienna, Austria). GAW and JDL acknowledge Discovery Grant support from the Natural Sciences and Engineering Research Council of Canada (NSERC). II acknowledges support from Bulgarian NSF grants D002-85 and D002-362.

\end{acknowledgements}


\end{document}